\documentclass[amssymb,fleqn,aps,twocolumn]{revtex4}
\usepackage[]{graphicx}
\usepackage[]{amsmath}
\usepackage{setspace}
\usepackage{hyperref}
\usepackage{setspace}
\usepackage{latexsym}

\def\n{\nu}
\def\a{\alpha}
\def\b{\beta}
\def\m{\mu}
\def\eps{\epsilon}

\begin{document}
\title{Ghost-free, finite, fourth order $D=3$ (alas) gravity}
\author{S Deser}
\email{deser@brandeis.edu}
\affiliation{Physics Department, Brandeis University, Waltham MA 02454 and \\
Lauritsen Laboratory, California Institute of Technology,  Pasadena CA 91125}
\date{\today}

\begin{abstract}

Canonical analysis of a recently proposed [1] linear+quadratic curvature gravity model 
in $D=3$ establishes its pure, irreducibly fourth derivative, quadratic curvature limit as
both ghost-free and power-counting UV finite, thereby maximally violating standard folklore. 
This limit is representative of a generic class whose kinetic terms are conformally invariant in any dimension, 
but it is unique in simultaneously avoiding the transverse-traceless graviton ghosts plaguing 
$D>3$ quadratic actions as well as double pole propagators in its other variables. While the 
two-term model is also unitary, its additional mode's second derivative nature forfeits finiteness. 

\end{abstract}
\maketitle

It is a truism of Lorentz-invariant local field theory that fourth (or higher) derivative actions entail ghosts: Any Lagrangian of the form 
$L = XO(M)O(m)X$, with $O(m)$ a (tensorial) Klein-Gordon operator of mass $m$ (possibly 0), and $X$ a (tensorial) field, has ghost propagator
$P^{-1} \approx 1/O(M) - 1/O(m)$. Of the apparent exceptions, Gauss-Bonnet-Lovelock gravity and scalar "Galileons" [2] have neither kinetic terms 
nor higher derivatives. Pure scalar curvature, $L= R + R^2$ models, do have higher derivatives at metric level, but are merely second-order in their proper, scalar-tensor, incarnations, a point whose analog we shall encounter here. Quadratic curvature models with torsion, but with affinities as independent variables, are thereby not strictly higher order. We also exclude prescriptions that simply "improve" the signs of ghost poles by fiat. 

For background, ghosts--excitations with negative probability (in quantum language), are unacceptable physically, as their existence destabilizes a system, much as do their classical counterparts, negative energy excitations, by destroying the ground state. Historically, they have nevertheless continued to exert interest in quantum field theory: 
Fourth order kinetic energies mitigate the ultraviolet catastrophes of loop corrections, most notably in $D=4$ General Relativity [3] whose coupling constant has inverse length dimensions, requiring an infinite number of additional terms that destroy its predictive power. I emphasize that the present $D=3$ toy model does NOT (the ``alas" in the title) indicate how to get physical relief for real, $D=4$ gravity: its very modest point is that even the venerable linkage between higher derivative actions and ghosts/negative energy modes is not airtight. 

This folk theorem has remained unchallenged until a recent [1] ($D=3$) linear plus quadratic curvature model claimed, by first reparameterizing the two-term metric action into a "two-tensor" form, to really represent two massive ghost-free spin-2 modes,
governed by the, second order, Fierz-Pauli action.  One appealing way to motivate this, at first sight unlikely, ghost-avoidance 
is as follows. If the second-order equation $O(m)X = 0$ permits only a vacuum solution, then the corresponding $1/O(m)$ ``propagator''
does not propagate any excitations; then the effective derivative order of $O(M)$ $[O(m)X] = 0$, or $O(m) [O(M)X]=0$, drops from 4 to 2. 
But the (linearized in $h = g- \eta$) Einstein tensor $G(h)=O(0)h$, being the full Riemann curvature in $D = 3$, is the perfect (and unique, 
as we shall see) exemplar of this mechanism: its vanishing implies flat space in (and only in) $D = 3$, where its propagator is pole-free. Furthermore, the specific proposed quadratic combination ensures that ``$O(M)$'' is the correct, separately ghost-free, FP operator. 
The pure quadratic case is more like $O(0)^2 X=0$, and as we shall see, it is not amenable to the reparameterization of [1]; whether 
it is ghost-free therefore requires detailed, metric, study. This is our main purpose: we will conclude that our limiting model indeed violates    
the folk theorem, a first for an intrinsically fourth derivative action. It is perhaps worth emphasizing at the outset that no miraculous derivative evaporation occurs: the field equations remain of fourth order, but two derivatives form a (harmless) Laplacian rather than a d'Alembertian, by the workings of $D=3$ tensor dynamics.

The analysis will be performed entirely in metric terms, in a linearized expansion about flat space, where the degree of freedom content can be analyzed without the irrelevant complications of nonlinearity.The parameterization proposed in [1] is treated in the Appendix. 
[Separately, it is an old story that any model involving polynomials in curvature allows $(A)dS$, as well as flat, vacua even without an explicit cosmological term (see, {\it e.g.}, [4]); these states, also treated in [1], are not relevant in the present context.]  The canonical decomposition, much simpler in $D = 3$ than in $D = 4$, will be further simplified by use of (linearized) gauge invariance. The Lagrangians of [1] are a one-parameter class,
\begin{eqnarray}
   I[h]&=& \int d^3x L(h) \\
   &=& \int d^3x \{-\frac{1}{2} m^2 R + (G^{\m\n} G_{\m\n} - 1/2 (tr G )^2)\}, \nonumber 
\end{eqnarray}
up to an overall, dimension $L^1$, gravitational constant that is set to unity. Our main focus is on the special, $m=0$, limiting case. 
It is actually part of a class of (pseudo)conformal-invariant actions 
whose linearizations are invariant in any $D$, while their full extensions scale 
as powers of the conformal factor. Its $D$=4 representative is just the familiar
Weyl action, which scales as the zeroth power. These actions are
\begin{eqnarray}
    I&=&\int d^Dx S_{\m\n} S_{\a\b} (g^{\m\a} g^{\n\b} - g^{\m\n} g^{\a\b}){\sqrt{-g}}, \nonumber\\
    S_{\m\n} &\equiv& R_{\m\n} - 1/2(D-1)\  g_{\m\n} R  
\end{eqnarray}
$S_{\m\n}$ is the Schouten tensor. Its curl, in $D$=3, is the familiar Cotton-Weyl 
tensor, initially introduced in $D$=4 Einstein gravity [5].

We decompose the metric deviation's components $h_{\m\n}$ into their orthogonal parts, insert these into the curvature and form the various scalars in (1) to display the action in terms of the independent, gauge-invariant, metric variables, where its excitation content becomes manifest. Conventions are $\eps^{0ij} = +1$, signature ($ - + +$); the Einstein tensor is defined by (5) below. The 2+1 decomposition of the (six) $h_{\m\n}$ is
\begin{eqnarray}
h_{ij} &=& (\partial_i h_j + \partial_j h_i) +\epsilon^{il} \epsilon^{jk} \phi_{lk}, \nonumber\\
h_{0i} &=& \eta_i + \epsilon^{ij} \psi_j,\,\,\,        h_{00}=n;
\end{eqnarray}
subscripts on the (indexless) variables denote normalized spatial derivatives, $\partial_i/ \sqrt{-\nabla^2}$, to keep standard dimensions for h and G. This decomposition is just the degenerate limit of the 
usual orthogonal one of [5], valid in all $D>3$,
\begin{eqnarray}
h_{ij} &=& h_{ij}^{TT} + (\partial_i h_j + \partial_j h_i) +\epsilon^{ilab..} \epsilon^{jkab..} \phi_{lk}, \nonumber\\
h_{0i} &=& \eta_i +  \partial_j \psi_{ij},\,\,\,        h_{00}=n;          
\end{eqnarray}
where $\psi_{ij}$ is an antisymmetric tensor that reduces to a scalar in 2-space. The crucial difference between (3) and (4) is that the familiar transverse-traceless,
$\partial_i h_{ij}^{TT}=0=h_{ii}^{TT}$, ``graviton" variable is identically zero in 2-space. This preserves 
the model from the ghosts in the term $L\sim h^{TT} \Box^2 h^{TT}$ that plague quadratic actions in 
$D>3$. The second, bigger, surprise that emerges below is that there at all exists a 
quadratic, 4th derivative covariant action whose (non-TT) variables avoid double poles!

Turning to our canonical analysis, gauge invariance of the action lets us set the 
three gauge parts $h_i$ and $\eta$ of the metric to zero by imposing the usual gauge choice $h_{ij,j} = 0 = h_{0i,i}$. There remain only the three gauge-invariant components ($\phi,\psi, n$) in (3). The Einstein tensor,
\begin{equation}
    G^{\m\n} = \frac{1}{2}\eps^{\m\alpha\beta} \eps^{\nu\lambda\sigma} \partial_\beta \partial_\sigma h_{\alpha\lambda},                               
\end{equation}
is easily verified to have the following 2+1 components,
\begin{eqnarray}
&2& G^{00}= - \nabla^2 \phi,\,\,\,    2 G^{0i}=  {\sqrt{-\nabla^2}\dot{\phi}_i + \eps^{il} (-\nabla^2)  \psi_l},\\
&2& G^{ij}=\ddot{\phi} _{ij} +\eps^{ip} \eps^{jq} (-\nabla^2) n_{pq} + (\eps^{ik} \sqrt{-\nabla^2} \dot{\psi}_{kj} + i\leftrightarrow j ).\nonumber
\end{eqnarray}

As a check, (6) manifestly obeys the Bianchi identity $G^{\m\n},_\n=0$. Orthogonality of the various ($h, G$) components under integration then easily yields the canonical form of the action (1),
\begin{eqnarray}
I &=& \int d^3x \{\frac{1}{2} \psi(- \nabla^2) (\Box- m^2) \psi +\frac{1}{2}  m^2 \phi (\Box - m^2)\phi\nonumber\\ 
&+&\frac{1}{8}  [\nabla^2 n - (\Box-2m^2) \phi]^2\}. 
\end{eqnarray}

Consider first the pure quadratic $m=0$, fourth order, action. It is the sum of a (non-ghost: $-\nabla^2$ is positive) massless mode plus an 
irrelevant complete square: the no-go theorem is successfully violated! There is one nondynamical relation, 
$\nabla^2 n- \Box \phi=0$, between $\phi$ and $n$, due to the Weyl/conformal invariance exhibited in (2). In the two-term massive branch, the Einstein term not only adds a (correct sign) $m^2$ to the d'Alembertian acting on the vector, 
but the previous pure multiplier part is no longer a perfect square, restoration of which reveals an additional $\phi$-mode described by the second term in (7). Unlike the first, vectorial one, it is a (spatial) massive tensor. It is worth noting that changing the relative coefficients in the quadratic combination (so losing its Weyl invariance), by adding $\delta L \approx R^2$ to (1) destroys the good properties of both branches, as one might expect from the preferred status of just this combination both as conformal invariant and as the special FP mass term respectively.
To summarize, the $m = 0$, pure fourth order, conformally invariant limit of (1) indeed successfully breaks the no-go 4th derivative theorem. 
[While its massive incarnation is likewise physical, as shown in the Appendix, this is because it is really a second-derivative two-field system, like the scalar-tensor form of $L = (R+R^2)$.]
Furthermore, the theory becomes power-counting finite: each higher loop adds a factor $\approx d^3k V^2 P^3$, where $V\approx k^4$ and $P\approx k^{-4}$ are respectively the vertex and propagator. Hence there is a net gain of one power of $1/k$, overcoming the one-loop (formally cubic) divergence by (at latest) 5-loop order.  [To avoid misunderstanding as to power counting, note that while one may 
remove two derivatives from the free action by a simple field redefinition that absorbs 
a factor $\sqrt{-\nabla^2}$ in each $\psi$, this would NOT change the net UV counting because 
each $\psi$ in the vertices would acquire the inverse of this factor.]
Also, there are no conformal anomalies in odd $D$. However, given the theory's special nature, perhaps not too much should be read into this first viable quantum gravity! In the above context, we resolve the seeming paradox that the ($m^2 R+R^2$) model seems to be both 4th, and (in its FP version) 2nd, order. The answer is clear from (7): Only the vector's propagator behaves as $1/k^4$; the tensor's just goes as $1/k^2$. Hence 
massive theory is nonrenormalizable, as expected also from the bad, 1/L, dimension of its Einstein term.

Some final comments: (A) The, third derivative order, fermionic SUGRA extensions of the above tensor model are its vector-spinor 
companions: the $D = 3$ Rarita-Schwinger equation just states that the vector-spinor field strength $f^\m= \eps^{\m\alpha\beta} D_\alpha \psi_\beta$ vanishes; there are no excitations,
so there is {\it a priori} hope of evading the no-go theorem here too, by adding the equivalents of the quadratic curvature terms $fDf$. The results will mirror our bosonic outcomes [respectively  for pure $L\approx f D f$ and $L\approx(f D f+m^2 \psi f)$], as guaranteed by SUSY. (B) The present ($m = 0$) miracle fails both in $D > 3$ where Riemann becomes 4-index, and in $D = 2$, where it has none; as usual, $D = 3$ is special. (C) Vectors are not viable candidates: their ``Riemann tensor" $F = curl A$ is only of first derivative order: correspondingly, the higher derivative models are only third ($F\partial F$) order, hence not ghost-prone [6] in the first place. (D) I have not studied extensions of the present model through addition of Chern-Simons (also conformally invariant), cosmological, or explicit mass terms, nor included (necessarily traceless) sources. 

I thank Paul Townsend for a conversation at the Imperial College Duffest where this work was begun, for later informing me that O. Hohm 
had also noted the "motivational", $G(h)\leftrightarrow O(0)X$, argument in text and for subsequently insisting that since massive FP exorcizes ghosts, they must also disappear (as indeed they finally did) from the massive metric form. This work was supported by Grants NSF 07-57190 and DOE DE-FG02-92-ER40701.

\appendix*
\section{}

In this Appendix I review the proposal of [1] relating their fourth order gravity model (1), to the ``two-field", 
second derivative theory at linear order:
\begin{equation}
L= (f-1/2 h)_{\m\n} G^{\m\n}(h) - \frac{1}{4}m^2 [(f_{\m\n})^2- (tr f)^2].          
\end{equation}    
in terms of present conventions. First, we show how simple completion of the square in (A.1) 
exhibits the respective single field actions. Decomposing the tensors into their traceless parts $(f'_{\m\n},G'_{\m\n})$ and their traces (f,G),
\begin{eqnarray}
L(f,h)&=&  -\frac{1}{4}m^2 [ f'_{\m\n} - 2m^{-2} G'_{\m\n}(h)]^2\nonumber\\
 &&+   \frac{1}{6}m^2 [f-\frac{1}{2}m^{-2}R^2 \nonumber\\
&& + m^{-2} [G_{\m\n}^2 -\frac{1}{2}G^2] - \frac{1}{2} hG(h),       
\end{eqnarray}
Dropping the irrelevant perfect squares, we recover (1), up to a trivial $m^2$ rescaling. On the other hand, 
we may also combine the terms of (A.1) as
\begin{equation}
L(f,h)= -\frac{1}{2} [(h-f) G (h-f)] + \frac{1}{2} \{f G(f) - \frac{1}{2}m^2 (f_{\m\n}^2\ -f^2)\},   
\end{equation}
where $G$ is the linearized Einstein operator (5).  The first term is again irrelevant, stating that $h-f=0$ up to gauge, while the rest is just the 
standard pure Fierz-Pauli action for f. However, the above procedures are valid only for
$m \neq 0$; equivalence is lost at $m=0$. Indeed, (A.1) states that both fields become trivial 
there: $G(h)=0=G(f)$, whereas (7) displays a perfectly physical massless mode. An apparent way around this has been suggested in the second paper of [1], in terms of 
a different initial form, which seems to yield an effective, second-order, Maxwell action for the 
massless case. Unfortunately, that procedure involves insertion of on-shell information into the
action (specifically inserting the solution of the linear Einstein equation-that its metric is pure gauge-
into the remaining terms), which is of course not permitted. We conclude that the pure quadratic, 
m=0, theory is irreducibly 4th order, without the 2nd order avatar underlying the massive case.
Hence, as explained in text, only it is the novel exception.

\end{document}